\documentclass[aps,prd,twocolumn,nofootinbib,showpacs]{revtex4}

\usepackage{amssymb,amsmath}
\usepackage{graphicx,color}
\usepackage[colorlinks=true, linkcolor=blue, citecolor=blue, urlcolor=blue]{hyperref}

\def\be{\begin{equation}}
\def\ee{\end{equation}}
\def\bea{\begin{eqnarray}}
\def\eea{\end{eqnarray}}

\newcommand{\mbf}[1]{\mathbf{#1}}
\newcommand{\half}{\textstyle \frac{1}{2}}
 
\renewcommand{\bar}[1]{\overline{#1}}

\begin{document}

\title{Light-Front Holography and Supersymmetric Conformal Algebra:
A Novel Approach to Hadron Spectroscopy, Structure, and Dynamics}

\author{Stanley J. Brodsky$^{1}$, Guy F. de T\'eramond$^{2}$, 
Hans G\"{u}nter Dosch$^{3}$}
\affiliation{$^{1}$\mbox{SLAC National Accelerator Laboratory, Stanford University, Stanford, CA 94309, USA}
$^{2}$\mbox{Laboratorio de F\'isica Te\'orica y Computacional, Universidad de Costa Rica, 11501 San Jos\'e, Costa Rica}
$^{3}$\mbox{Institut f\"{u}r Theoretische Physik der Universit\"{a}t,  D-69120 Heidelberg, Germany}
}

\begin{abstract}

We give an overview of recent progress into the infrared structure of QCD based on the gauge/gravity correspondence and light-front quantization, where the color confining interaction for mesons and baryons is determined by an underlying superconformal algebraic structure. This new approach to hadron physics gives remarkable connections and predictions across the entire mass spectrum of hadrons and also describes the infrared behavior of the strong coupling. More recently, an extensive study of form factors, polarized and unpolarized parton distributions and the sea quark contribution to the nucleon has been carried out by extending the holographic formalism to incorporate the nonperturbative structure of Veneziano amplitudes. Contribution to the report ``Strong QCD from Hadron Structure Experiments” Jefferson Lab Workshop, 5-9 November, 2019.

\end{abstract}

\pacs{13.66.Bc, 13.66.De, 12.38.Bx}

\date{\today}



\maketitle

Recent  insights into the nonperturbative structure of QCD based on the gauge/gravity correspondence and light-front (LF) quantization, {\it light-front holography} for short~\cite{deTeramond:2008ht},  have lead to effective semiclassical bound state equations for mesons and baryons where the confinement potential is determined by an underlying superconformal algebraic structure~\cite{Brodsky:2013ar, deTeramond:2014asa, Dosch:2015nwa}. The formalism provides a remarkably  first approximation to QCD, including its hidden supersymmetric hadronic features. The resulting light-front  wave equation  allows the familiar tools and insights of Schr\"odinger's nonrelativistic quantum mechanics and the Hamiltonian formalism to be applied to relativistic hadronic physics~\cite{Brodsky:1997de, Brodsky:2014yha, Zou:2018eam}.  It should be noted that supersymmetry in this approach is supersymmetric quantum mechanics~\cite{Witten:1981nf} and refers to bound state wave functions and not to elementary quantum fields.

Our work in this area can be traced back to the original article of Polchinski and  Strassler~\cite{Polchinski:2001tt}, where the exclusive hard-scattering counting rules~\cite{Brodsky:1973kr, Matveev:ra} were derived from the warped geometry of Maldacena's five-dimensional anti-de Sitter AdS$_5$ space: The  hadron in elastic scattering at high momentum transfer shrinks to a small size near the AdS boundary at $z = 0$ where the dual space is conformal ($z$ is the fifth coordinate of  AdS space). Hadron form factors (FFs) look very different in AdS space~\cite{Polchinski:2002jw} or in physical spacetime~\cite{Drell:1969km,West:1970av}:  One can show, however, that a precise mapping can be carried out at Dirac's fixed light-front time~\cite{Dirac:1949cp} for an arbitrary number of partons~\cite{Brodsky:2006uqa}. As a result, the impact parameter generalized parton distributions~\cite{Soper:1976jc, Burkardt:2000za}  are expressed in terms of the square of AdS  eigenmodes, provided that the invariant transverse impact variable $\zeta$ for the $n$-parton bound state is identified with the holographic variable $z$. For a two-parton system, $\zeta^2 = x(1-x)  \mbf{b}^2_\perp$, the AdS modes are mapped directly to the square of effective light-front wave functions (LFWFs) which incorporate the nonperturbative pole structure of FFs~\cite{Brodsky:2006uqa}.  Similar results follow from the mapping of the matrix elements of the energy-momentum tensor~\cite{Brodsky:2008pf}.

A semiclassical approximation to light-front QCD follows from the LF Hamiltonian equation  $P_\mu P^\mu  \vert \psi \rangle = M^2 \vert \psi \rangle$  with  $P = \left(P^-, P^+, \mbf{P}_\perp\right)$.   In the limit  $m_q \to 0$ the LF Hamiltonian for a $q \bar q$ bound state can be systematically reduced to a wave equation in the variable $\zeta$~\cite{deTeramond:2008ht}
\be
\left(-\frac{d^2}{d\zeta^2} 
- \frac{1 - 4L^2}{4\zeta^2}+ U(\zeta) \right)  \phi(\zeta) = M^2 \phi(\zeta) ,
\ee
where  the effective potential $U$ includes all interactions,  including those from higher Fock states. The orbital angular momentum $L = 0$ corresponds to the lowest possible solution. The LF equation has similar structure of wave equations in AdS, and can be embedded in AdS space provided that $\zeta = z$~\cite{deTeramond:2008ht}. The precise mapping allows us to write the LF confinement potential $U$ in terms of the dilaton profile which modifies AdS~\cite{deTeramond:2010ge}.

The separation of kinematic and dynamic components can be extended to arbitrary integer-spin $J$  by starting from a dilaton-modified AdS action for a rank-$J$ symmetric tensor field and  $\Phi_{N_1 \dots N_J}$. Variation of the AdS action leads to a general wave equation plus kinematical constraints to eliminate lower spin from the symmetric tensor~\cite{{deTeramond:2013it}}. LF mapping  allows to determine the mass function in the AdS action in terms of physical kinematic quantities consistent with the AdS stability bound~\cite{Breitenlohner:1982jf}. Similar derivation for arbitrary half-integral spin follows for Rarita-Schwinger  spinors in AdS~\cite{deTeramond:2013it}. In this case, however, the  dilaton term does not lead to an interaction~\cite{Kirsch:2006he}  and an effective Yukawa-type interaction has to be introduced instead~\cite{Abidin:2009hr}.  Embedding light-front physics in a higher dimension gravity theory leads to important insights into the nonperturbative structure of bound state equations in QCD for arbitrary spin, but does not answer how the effective confinement dynamics  is determined and how it can be related to the symmetries of QCD itself?

\begin{figure}[h]
\begin{center}
\includegraphics[width=8.6cm]{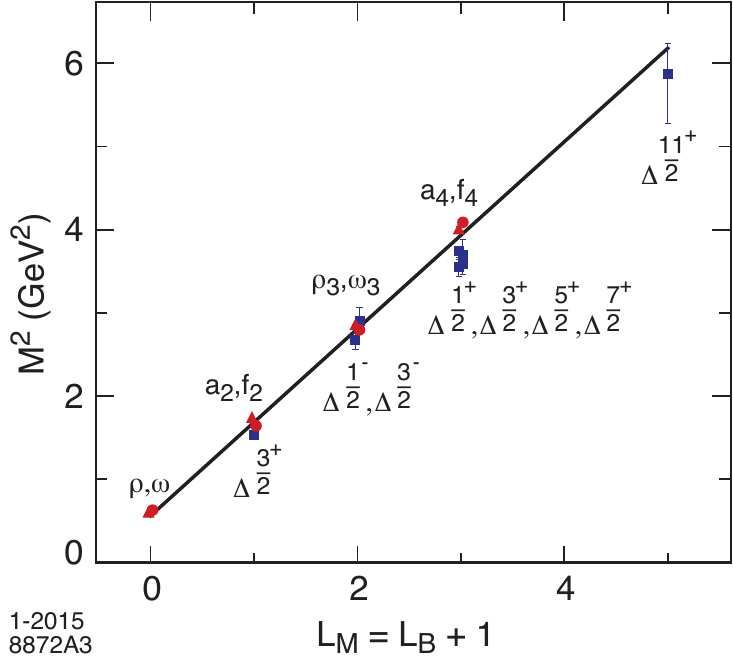}
\end{center}
\caption{\label{rho-delta}  Supersymmetric vector meson and $\Delta$  partners from Ref.~\cite{Dosch:2015nwa}  The experimental values of $M^2$ for confirmed states~\cite{Tanabashi:2018oca}  are plotted vs $L_M = L_B+1$. The solid line corresponds to  $\sqrt \lambda= 0.53$ GeV. The $\rho$ and $\omega$ mesons have no baryonic partner, since it would imply a negative value of $L_B$.}
\end{figure}

Conformal algebra underlies in LF holography the scale invariance of the QCD Lagrangian~\cite{Brodsky:2013ar}. It leads to the introduction of a scale $\lambda = \kappa^2$ and harmonic confinement, $U \sim \lambda \zeta^2$,  maintaining the action conformal invariant~\cite{Brodsky:2013ar, deAlfaro:1976je}. The oscillator potential corresponds to a dilaton profile and thus to linear Regge trajectories~\cite{Karch:2006pv}. Extension to superconformal algebra leads to a specific connection between mesons and baryons~\cite{Dosch:2015nwa} underlying the $SU(3)_C$ representation properties, since a diquark cluster can be in the same color representation as an antiquark, namely $\bar 3 \in 3 \times 3$.   We follow~\cite{Fubini:1984hf} and define the fermionic generator $R_\lambda =  Q + \lambda S$ with anticommutation relations $\{R_\lambda,R_\lambda\} =   \{R_\lambda^\dagger, R_\lambda^\dagger\} = 0$. It generates a new Hamiltonian   $G_\lambda = \{R_\lambda, R_\lambda^\dagger\}$ which closes under the graded algebra  $ [R_\lambda, G_\lambda]  = [R_\lambda^\dagger, G_\lambda] = 0$.   The generators $Q$ and $S$ are related to the generator of time translation  $H = \half \{Q,Q^\dagger\}$~\cite{Witten:1981nf}  and special conformal transformations   $K = \half\{S,S^\dagger\}$: together with the generator of dilations $D$ they satisfy the conformal algebra. The new Hamiltonian $G_\lambda$ is an element of the superconformal (graded) algebra and uniquely determines the bound-state equations for both mesons and baryons~\cite{deTeramond:2014asa,Dosch:2015nwa}
\bea
 \left(-\frac{d^2}{d\zeta^2} + \frac{4 L_M^2 -1}{4 \zeta^2}  +  V_M(\zeta) \right)\phi_M &=&  M^2 \, \phi_M,  \label{M}\\
\left(-\frac{d^2}{d\zeta^2} + \, \frac{4 L_B^2 -1}{4 \zeta^2}  + V_B(\zeta)  \right)\phi_B &=&  M^2 \, \phi_B , \label{B}
\eea
including essential constant terms in the effective confinement potential  $ V_{M,B}(\zeta) = \lambda_{M,B}^2\, \zeta^2 + 2 \lambda_{M,B} (L_{M,B} \mp 1)$,  with $\lambda_M = \lambda_B \equiv \lambda$ (equality of Regge slopes) and  $L_M = L_B + 1$~\cite{Note1}. This is shown in Fig.~\ref{rho-delta}.   The mass spectrum from (\ref{M}-\ref{B}) is $M^2_M = 4 \lambda (n+ L_M )$ and $M^2_B = 4 \kappa^2 (n+ L_B+1)$ with the same slope in $L$ and $n$, the radial quantum number.  Since $[R_\lambda^\dagger, G_\lambda] = 0$, it follows that the state  $\vert M, L  \rangle$ and $R_\lambda^\dagger \vert M, L \rangle = \vert B, L - 1\rangle$ have identical eigenvalues $M^2$, thus  $R_\lambda^\dagger$ is interpreted as the transformation operator of a single constituent antiquark (quark) into a diquark cluster with quarks (antiquarks) in the conjugate color representation. The pion, however, has a special role as the unique state of zero mass which is annihilated by $R_\lambda^\dagger$,  $R_\lambda^\dagger \vert M, L = 0 \rangle = 0$:  The pion has not a baryon partner and thus breaks the supersymmetry.

\begin{figure}[h]
\begin{center}
\includegraphics[width=8.6cm]{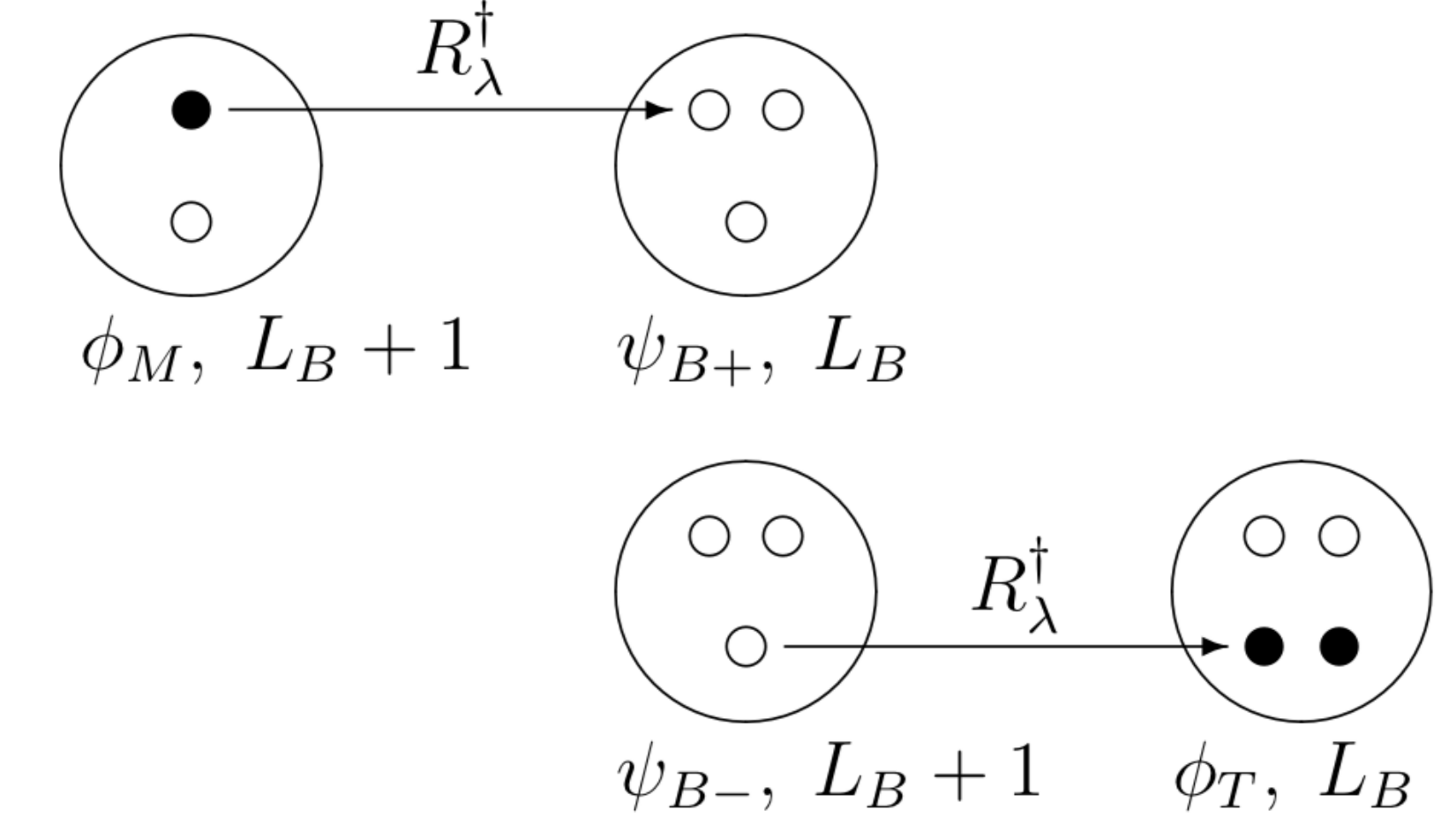}
\end{center}
\caption{\label{4-plet}  Supersymmetric 4-plet representation  of same-mass and parity hadronic states $\{\phi_M, \psi_{B+}, \psi_{B-},\phi_T\}$~\cite{Brodsky:2016yod}.  Mesons are interpreted as $ q \bar q$ bound states, baryons as quark-antidiquark bound states and  tetraquarks as diquark-antidiquark bound states. The fermion ladder operator $R^\dagger_\lambda $ connects antiquark (quark) and  diquark (anti-diquark) cluster of the same color.  The baryons have  two chirality components with orbital angular momentum  $L$ and $L+1$.}
\end{figure}

Embedding in AdS is also useful to extend the superconformal Hamiltonian to include the spin-spin interaction: From the spin dependence of mesons~\cite{deTeramond:2013it}  one concludes that $G_\lambda = \{R_\lambda, R_\lambda^\dagger\} + 2 \lambda s$, with $s = 0, 1$ the total internal spin of the meson or the spin of the diquark cluster of the baryon partner~\cite{Brodsky:2016yod}.  The lowest mass state of the vector meson family, the $\rho$ (or the $\omega$) is  also annihilated by the  operator $R^\dagger$, and has no baryon partner: The effect of the spin term  is an overall shift of the quadratic mass scale without a modification of the LFWF as depicted in Fig.~\ref{rho-delta}. The analysis  was consistently applied to the radial and orbital excitation spectra of the $\pi, \rho, K, K^*$ and $\phi$  meson families, as well as to the $N, \Delta, \Lambda, \Sigma, \Sigma^*, \Xi$ and $\Xi^*$ in the baryon sector, giving the value  $\kappa = \sqrt{\lambda} = 0.523 \pm 0.024 \, {\rm GeV}$ from the light hadron spectrum~\cite{Brodsky:2016yod}. Contribution of quark masses~\cite{Brodsky:2008pg} are included via the Feynman-Hellman theorem, $\Delta M^2 = \langle \sum_q m_q^2/x_q \rangle$, with the effective values $m_u = m_d = 46$ MeV and $m_s = 357$ MeV~\cite{Brodsky:2014yha}. The complete multiplet is obtained by applying the fermion operator $R_\lambda^\dagger$ to the negative-chirality component baryon wave function~\cite{deTeramond:2014asa, Brodsky:2014yha}  $\phi_B = \left\{\psi_{+}(L_B), \psi_{-}(L_B+ 1)\right\}$ leading to a tetraquark bosonic partner, $R_\lambda^\dagger\, \psi_{-}  =\phi_T$, a bound state of diquark and anti-diquark clusters with angular momentum $L_T=L_B$~\cite{Brodsky:2016yod}: The full supermultiplet (Fig.~\ref{4-plet}) contain mesons, baryons and tetraquarks~\cite{Note2}.  A systematic analysis of the isoscalar bosonic sector was also performed using the framework described here; the $\eta'-\eta$ mass difference is correctly reproduced~\cite{Zou:2019tpo}.

We have shown in~\cite{Dosch:2015bca} that the basic underlying hadronic supersymmetry still holds and gives remarkable connections across the entire spectrum of light and heavy-light hadrons even if quark masses break the conformal invariance. In particular, the $L = 0$ lowest mass meson defining the $K, K^*, \eta', \phi, D, D^*, D_s, B, B^*,B_s$ and $B^*_s$ families examined in~\cite{Dosch:2015bca} has in effect no baryon partner, conforming to the SUSY mechanism found for the light hadrons. The analysis was extended in~\cite{Dosch:2016zdv} by showing that the embedding of the light-front wave equations in AdS space nevertheless determines the form of the confining potential in the LF Hamiltonian to be harmonic, provided that: a) the longitudinal and transverse dynamics can be factored out to a first approximation and b)  the heavy quark mass dependence determines the increasing value of the Regge slope according to Heavy Quark Effective Theory (HQET)~\cite{Isgur:1991wq}. This model has been confronted with data in the detailed analysis performed in~\cite{Nielsen:2018uyn} including  tetraquarks with one charm or one bottom quark as illustrated in Tables~\ref{c} and \ref{b}. The double-heavy  hadronic spectrum, including mesons, baryons and tetraquarks and their connections was examined in~\cite{Nielsen:2018ytt} confirming the validity of the supersymmetric approach applied to this sector. The lowest mass meson of each family, the $\eta_c, J/\Psi, \eta_B$ and $Y$ have no hadronic partner and the increase in the Regge slope qualitatively agrees with the HQET prediction.

\begin{figure}[h]
\begin{center}
\includegraphics[width=7.8cm]{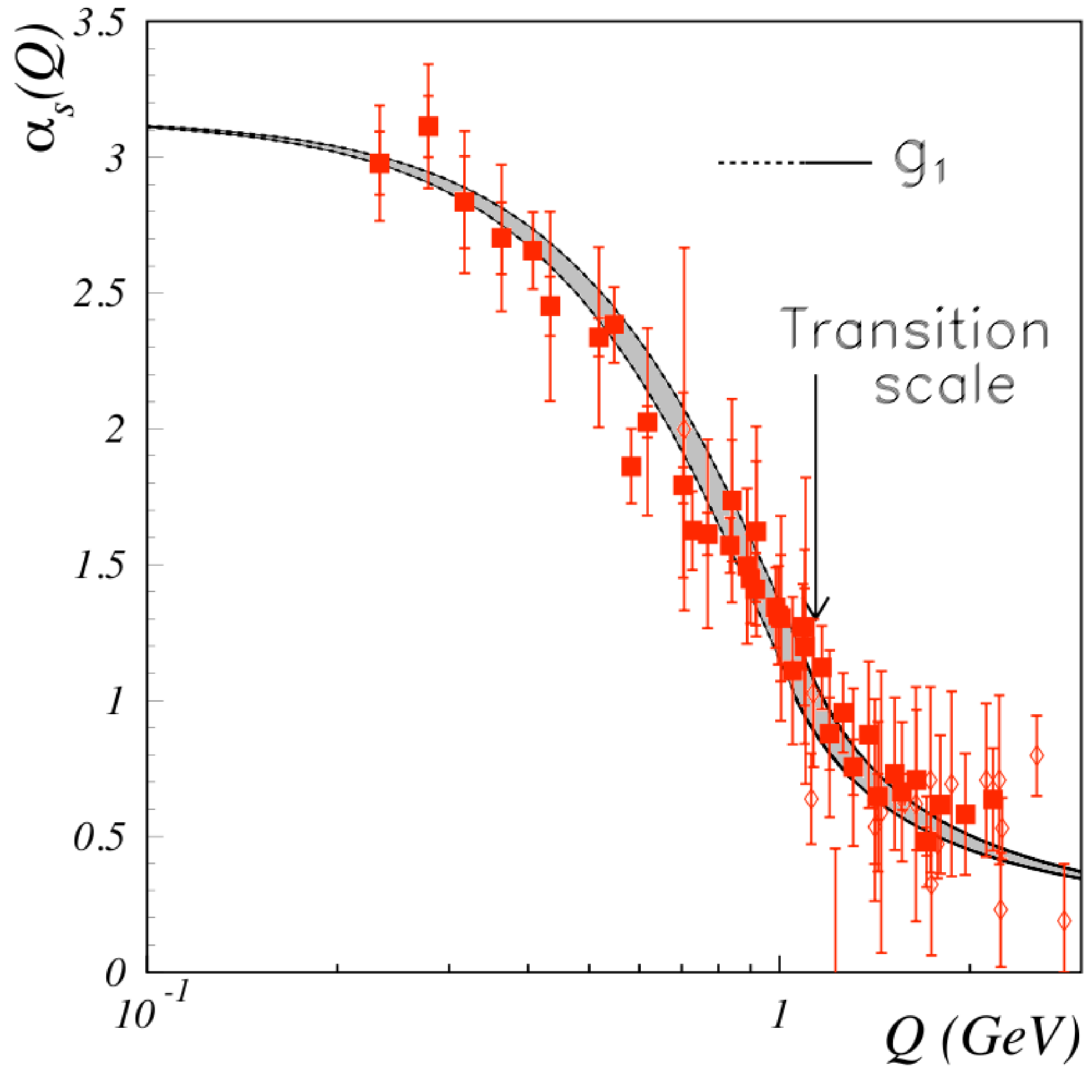}
\end{center}
\caption{\label{alphaIR} Matching the nonperturbative and perturbative couplings regimes at 5-loop $\beta$-function in the $\bar{MS}$ renormalization scheme and comparison with $\alpha_s$ measurements from the Bjorken sum rule.   For $\sqrt{\lambda} =  0.523 \pm 0.024$ GeV we obtain $\Lambda_{\bar{MS}}=0.339 \pm 0.019 ~  {\rm GeV}$ compared with the world average $\Lambda_{\bar{MS}}= 0.332 \pm 0.017$ GeV  \cite{Deur:2016opc}.}
\end{figure}

Embedding LF dynamics in AdS allow us to study  the infrared (IR) behavior of the strong coupling. In fact, it is possible to establish a connection between the short-distance behavior of the QCD coupling  $\alpha_s$ with JLab long-distance measurements of $\alpha_s$ from the Bjorken sum rule~\cite{Brodsky:2010ur, Deur:2014qfa, Deur:2016cxb, Deur:2016opc}. In light front holography the IR strong coupling is $\alpha_s^{IR}(Q^2) = \alpha_s^{IR}(0) e^{- Q^2/4 \lambda}$. One can obtain $\Lambda_{QCD}$  from matching the perturbative (5-loop) and nonperturbative couplings at the transition scale $Q_0$ as shown in Fig.~\ref{alphaIR}. For $\sqrt{\lambda} =  0.523 \pm 0.024$ GeV we find $\Lambda_{\bar{MS}}=0.339 \pm 0.019 ~  {\rm GeV}$ compared with the world average $\Lambda_{\bar{MS}}= 0.332 \pm 0.017$ GeV and $Q_0^2 \simeq 1$ GeV$^2$.  Therefore, one can establish a connection between the proton mass $M^2_p = 4 \lambda$ and the perturbative QCD scale $\Lambda_{QCD}$ in any renormalization scheme.

An extensive study of form factors (FFs)~\cite{Sufian:2016hwn} and parton distributions~\cite{deTeramond:2018ecg, Liu:2019vsn}  has been carried out recently using an extended model based on the gauge-gravity correspondence, light-front holography, and the generalized Veneziano model~\cite{Veneziano:1968yb,Ademollo:1969wd,Landshoff:1970ce}. The nonperturbative strange and charm sea content of the nucleon has been studied by also incorporating constraints from lattice QCD~\cite{Sufian:2018cpj, Sufian:2020coz}. Meson~\cite{Brodsky:2011xx} and nucleon transition form factors, such as the proton to Roper $N(1440)1/2^+$ transition, can also be described within the light-front holographic framework~\cite{deTeramond:2011qp, Ramalho:2017pyc} and extended to other nucleon transitions~\cite{Gutsche:2019yoo}, such as the transition to the $\Delta(1232)3/2^+,  N(1520)3/2^-,  N(1535)1/2^-,  \Delta(1600)3/2^+$ and $\Delta(1620)1/2^-$ states measured at CLAS~\cite{Mokeev:2015lda}.

Hadron FFs in the light-front holographic approach are  a sum from the Fock expansion of states
$ F(t) = \sum_\tau c_\tau F_\tau(t)$,
where the $c_\tau$ are spin-flavor coefficients and $F_\tau(t)$  has the Euler's  Beta form structure ~\cite{Veneziano:1968yb,Ademollo:1969wd,Landshoff:1970ce} 
 \be
 F_\tau(t) = \frac{1}{N_\tau} B\big(\tau-1,  1 - \alpha(t) \big),
 \ee
where $\alpha(t)$ is the Regge trajectory of the vector meson which couples to the quark current in the hadron. For twist $\tau= N$, the number of constituents in a Fock component, the FF is an $N-1$ product of poles 
\be
F_{\tau}( Q^2) = \frac{1}{\big(1 + \frac{Q^2}{M^2_{n=0}}\big) \big(1 + \frac{Q^2}{M^2_{n=1}} \big) \cdots \big(1 + \frac{Q^2}{M^2_{n =\tau - 2}} \big)}, 
\ee
located at
$ - Q^2 = M^2_n = (n + 1 - \alpha(0)) / \alpha'$, 
which generates the radial excitation spectrum of the exchanged  particles in the $t$-channel~\cite{Brodsky:2014yha, Zou:2018eam}. The trajectory $\alpha(t)$ can be computed  within the superconformal framework and its intercept $\alpha(0)$ incorporates  the quark masses~\cite{Sufian:2018cpj}.

\begin{figure}[h]
\begin{center}
\includegraphics[width=8.6cm]{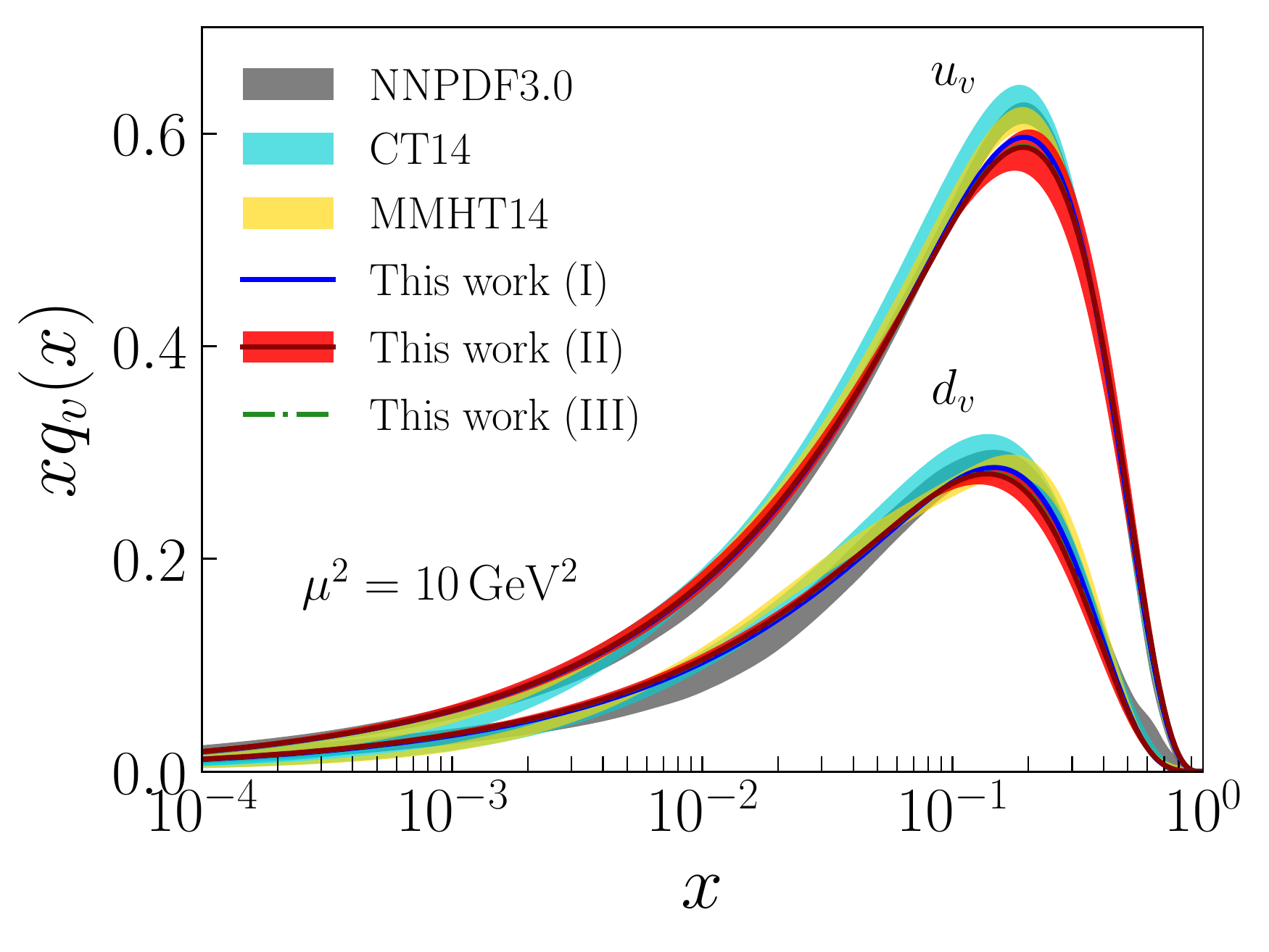}
\end{center}
\caption{\label{unpolPDFs} Comparison of $x q(x)$  in the proton from LF holographic QCD~\cite{deTeramond:2018ecg} with global fits~\cite{Ball:2014uwa, Harland-Lang:2014zoa, Dulat:2015mca} for models I, II and III in ~\cite{Liu:2019vsn}. The  results are evolved from the initial scale $\mu_0 = 1.06  \pm 0.15$ GeV~\cite{Deur:2016opc}.}
\end{figure}

 Using the integral representation of the Beta function the FF is expressed in a reparametrization invariant form 
 \be
 F(t)_\tau = \frac{1}{N_\tau} \int_0^1 dx w'(x) w(x)^{-\alpha(t)} \left[1 - w(x) \right]^{\tau -2} ,
 \ee
 with $w(0) = 0, \quad w(1) = 1, \quad   w'(x) \ge 0$. The flavor FF is given in terms of the valence GPD  at zero skewness
$F^q_\tau(t) = \int_0^1 dx \, q_\tau(x) \exp[t f(x)] $
with the profile function $f(x)$ and PDF $q(x)$ determined by $w(x)$
\bea
f(x)&=&\frac{1}{4\lambda}\log\Big(\frac{1}{w(x)}\Big) ,  \\ 
q_\tau(x)&=&\frac{1}{N_\tau}[1-w(x)]^{\tau-2}w(x)^{-\frac{1}{2}}w'(x)  .
\eea
Boundary conditions at $x \to 0$ follow from the expected Regge behavior, $w(x) \sim x$,  and  at $x \to 1$ from the inclusive-exclusive counting rules~\cite{Drell:1969km} $q_\tau(x) \sim (1-x)^{2 \tau - 3}$ which imply $w'(1) = 0 $. These physical conditions, together with the constraints written above, basically determine the form of $w(x)$. If the universal function $w(x)$ is fixed by the nucleon PDFs then the pion PDF is a prediction~\cite{deTeramond:2018ecg}. The unpolarized PDFs for the nucleon are compared with global fits in Fig.~\ref{unpolPDFs}.

\begin{figure}[h]
\begin{center}
\includegraphics[width=8.6cm]{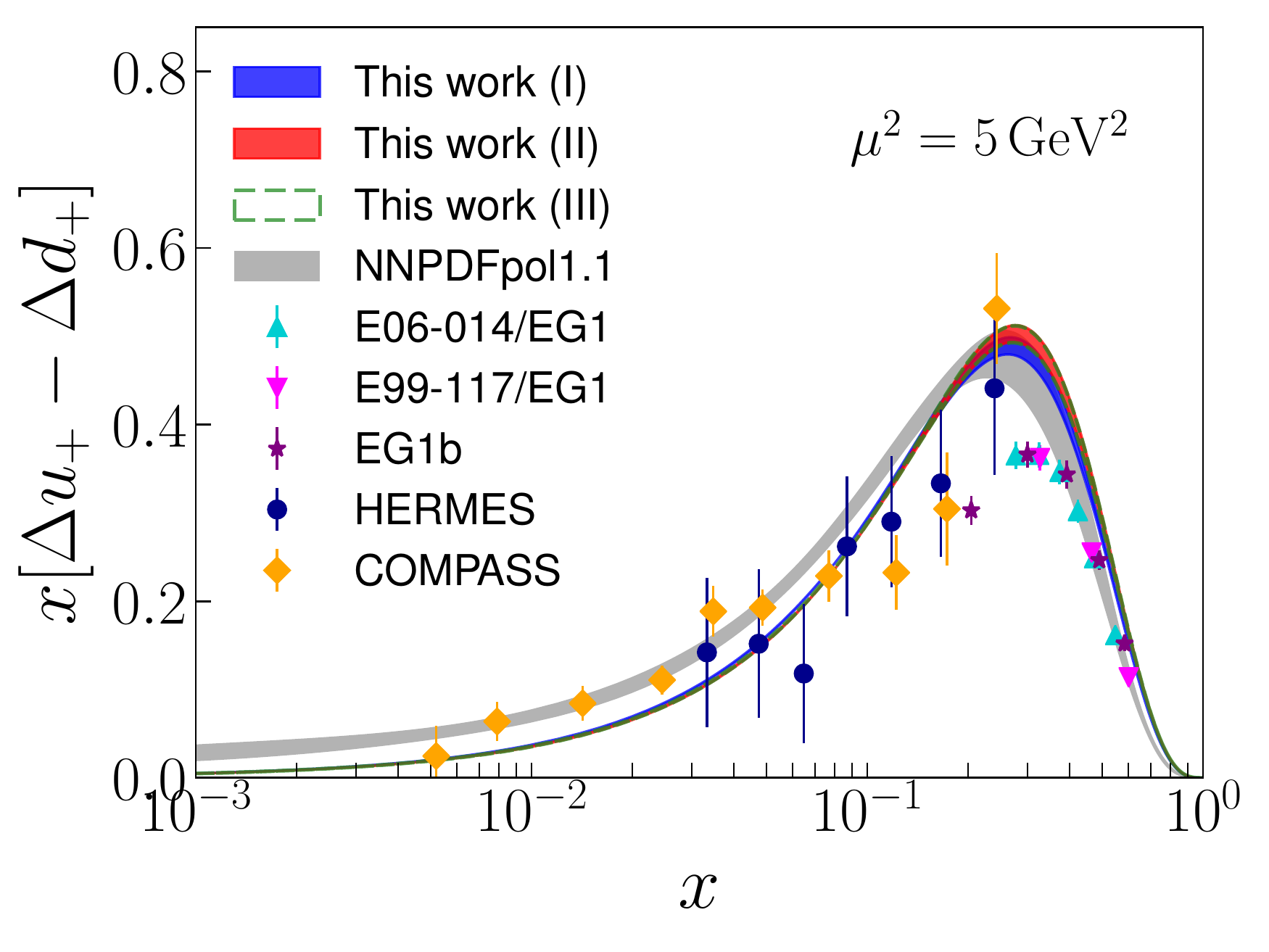}
\end{center}
\caption{\label{poldist} Polarized distributions of the isovector combination $x[\Delta u_+(x)-\Delta d_+(x)]$ in comparison with NNPDF global fit~\cite{Nocera:2014gqa} and experimental data~\cite{Zheng:2003un,Zheng:2004ce,Parno:2014xzb,Dharmawardane:2006zd,Airapetian:2003ct,Airapetian:2004zf,Alekseev:2010ub}. Three sets of parameters are determined from the Dirac form factor and unpolarized valence distributions.}
\end{figure}

\begin{figure}[h]
\begin{center}
\includegraphics[width=8.6cm]{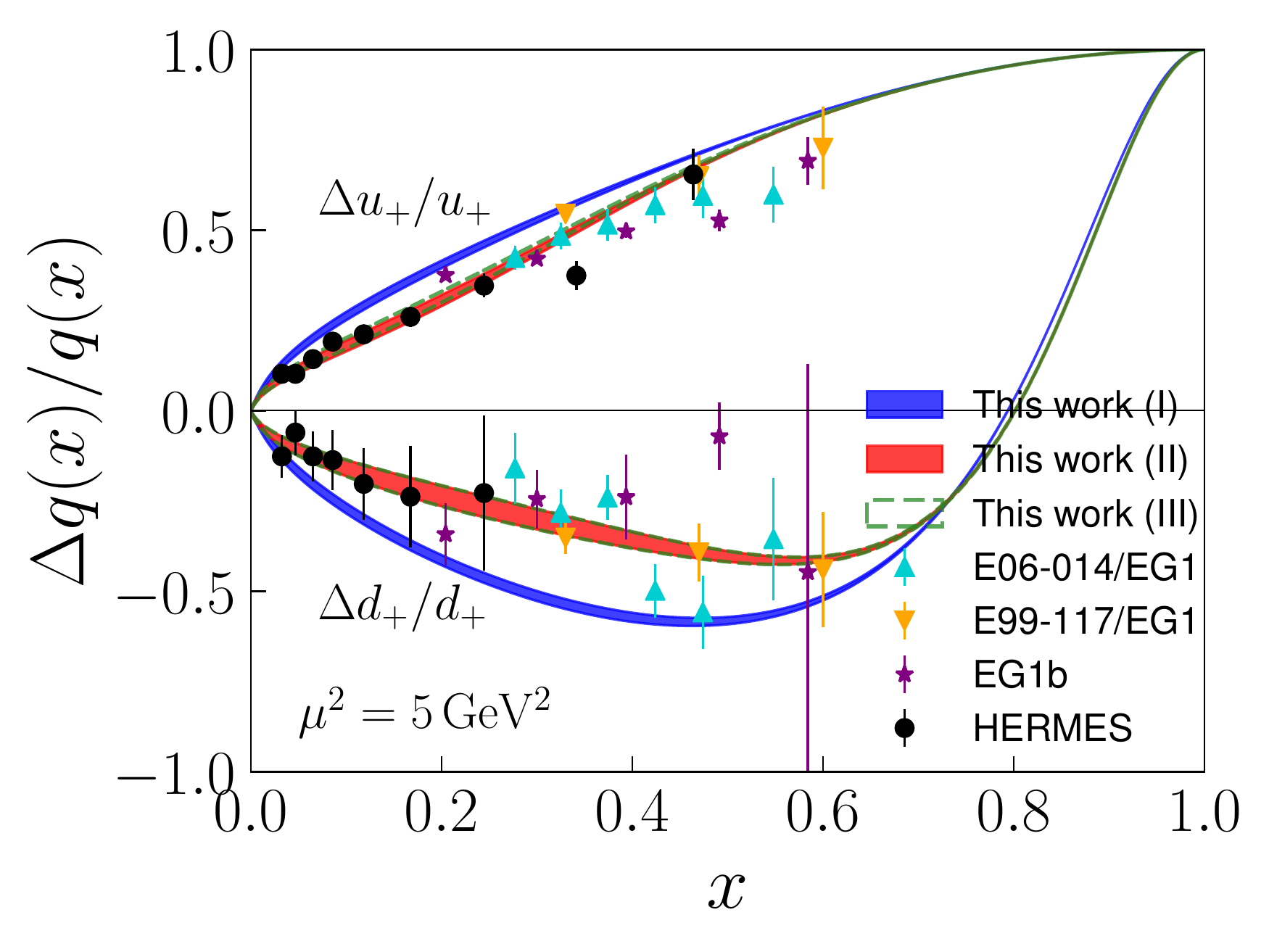}
\end{center}
\caption{\label{Dqoq}Helicity asymmetries of $u + \bar{u}$ and $d + \bar{d}$. Symbols as in Fig.~\ref{poldist}.}
\end{figure}

\begin{figure}[h]
\begin{center}
\includegraphics[width=8.6cm]{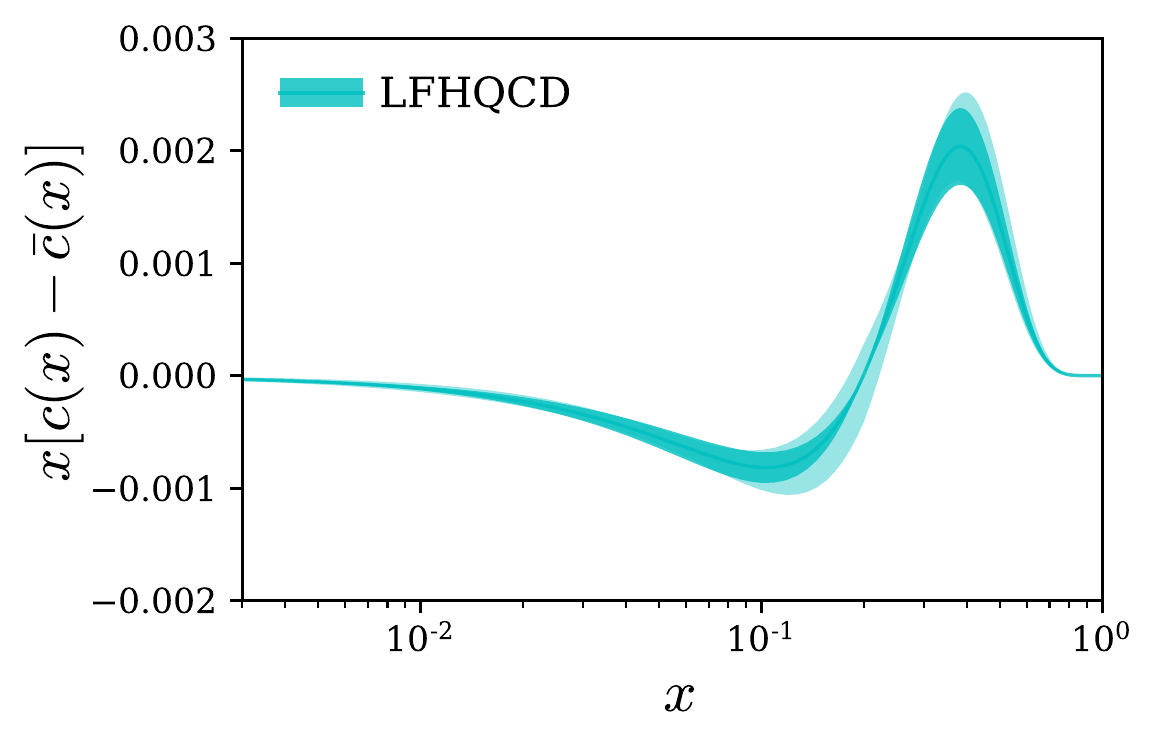}
\end{center}
\caption{\label{cbarc}The distribution function $x[c(x) - \bar{c}(x)]$ computed in the light front holographic framework using lattice QCD input of the charm electromagnetic form factors~\cite{Sufian:2020coz}.}
\end{figure}

To study the polarized GPDs and PDFs  we perform a  separation of chiralities in the AdS action: It allows the computation of  the matrix elements of the axial current –including the correct normalization, once the coefficients $c_\tau$ are fixed for the vector current~\cite{Liu:2019vsn}. The formalism incorporates the helicity retention between the leading quark al large $x$ and the parent hadron:
 $\lim_{x \to 1} ~  \frac{\Delta q(x)}{q(x)} = 1$, a  perturbative QCD result~\cite{Farrar:1975yb}. It also predicts no-spin correlation with the parent hadron at low $x$:
$\lim_{x \to 0} ~  \frac{\Delta q(x)}{q(x)} = 0$. We compare our predictions with available data for spin-dependent PDFs in Fig.~\ref{poldist} and for the ratio $\Delta q(x)/q(x)$ in Fig~\ref{Dqoq}.  The first lattice QCD computation of the the charm quark contribution to the electromagnetic form factors of the nucleon with three gauge ensembles (one at the physical pion mass) was performed in~\cite{Sufian:2020coz}. It gives the necessary constraints to compute the nonperturbative  intrinsic charm-anticharm asymmetry $c(x) - \bar c(x)$ using  the light front holography approach. The results are shown in Fig.~\ref{cbarc} ($q_+ \equiv q + \bar{q}$).

We have shown how the classical equations of motion for hadrons of arbitrary spin derived from the 5-dimensional gravity theory have the same form of the semiclassical bound-state equations for massless constituents in LF quantization. The implementation of superconformal algebra determines uniquely the form of the confining interaction. 
This new approach to hadron physics  incorporates basic nonperturbative properties which are not apparent from the chiral QCD Lagrangian, such as the emergence of a mass scale and the connection between mesons and baryons. In particular, the prediction of a massless pion in the chiral limit is a consequence of the superconformal algebraic structure and not of the Goldstone mechanism. The structural framework of LF holography also provides nontrivial connections between the structure of form factors
 and polarized and unpolarized quark distributions with nonperturbative results such as Regge theory and the Veneziano model.

Specific key results, such as the prediction of the ratio $\Delta q(x)/ q(x)$ at large $x$ will be tested very soon in upcoming experiments at JLab~\cite{E12-06-110,E12-06-122}. The strange-antistrange asymmetry could be  explored in semi-inclusive $\phi$ electroproduction with CLAS 12. Our study of the nucleon to Roper transition form factor will be extended  up to  $Q^2 = 12$ GeV for comparison with new CLAS data. The prediction of  hadron states within superconformal multiplets of meson-baryon-tetraquarks (for example the multiplets shown in Tables~\ref{c} and \ref{b})  can motivate the search for new tetraquark states.  Many  other important applications to hadron physics based on the holographic framework have been studied in addition to the new developments described here; unfortunately it is not possible to review them in this short overview and we apologize to the authors in advance.

Contribution to the report ``Strong QCD from Hadron Structure Experiments" Jefferson Lab Workshop, 5-9 November, 2019. We are grateful to Alexandre Deur, Tianbo Liu, Marina Nielsen,  Raza Sabbir Sufian and Liping Zou who have contributed greatly to the physics topics reviewed here. This research was partly supported by the Department of Energy,  contract DE--AC02--76SF00515.  
SLAC-PUB-17520.

\newpage

\section*{Figures}

\newpage

 \section*{Tables}

\begin{table*}[ht]
\caption{\label{c} Quantum number assignment of different meson families with quarks: $q=u,~d,~s$ and one charm quark $c$ and their supersymmetric baryon and tetraquark partners from Ref.~\cite{Nielsen:2018uyn}. Each family is separated by a horizontal line.  For  baryons multiplets with same $L_B$ and $S_D$  only the state with the highest possible value for $J$ is included. Diquarks clusters are represented by $[~]$ have total spin $S_D=0$, and  $(~)$ represents $S_D=1$.   The quantum numbers $J^P=1^+$ and $J^P=2^-$ are assigned  to the states $D(2550)$ and $D_J(2600)$, but their quantum numbers have not yet been determined. States with a question mark (?) are the predicted ones. The lowest meson bound state of each family has no baryon or tetraquark partner and breaks the supersymmetry.}
  \begin{ruledtabular}
  \begin{tabular}{ccc|ccc|ccc}
\multicolumn{3}{c|}{Meson} & \multicolumn{3}{c|}{Baryon} & \multicolumn{3}{c}{Tetraquark}\\
$q$-cont&$J^{P(C)}$ & Name & $q$-cont & $J^{P}$ &Name &$q$-cont &$J^{P(C)}$& Name \\
\hline
$\bar{q}c$&$0^{-}$&$D(1870)$& --- &---&---& --- & --- &--- \\
$\bar{q}c$&$1^{+}$&${D}_1(2420)$& $[ud]c$ & $(1/2)^+$&$\Lambda_c(2290)$ & $[ud][\bar{c}\bar{q}]$ & $0^{+}$ &$\bar{D}_0^*(2400)$ \\
$\bar{q}c$&$2^{-}$&$D_J(2600)$ & $[ud]c$ & $(3/2)^-$ & $\Lambda_c(2625)$ & $[ud][\bar{c}\bar{q}]$ & $1^{-}$ & --- \\
\hline
$\bar{c}q$&$0^{-}$&$\bar{D}(1870)$& --- &---&---& --- & --- &--- \\
$\bar{c}q$&$1^{+}$&$\bar{D}_1(2420)$& $[cq]q$ & $(1/2)^+$&$\Sigma_c(2455)$ & $[cq][\bar{u}\bar{d}]$ & $0^{+}$ &${D}_0^*(2400)$ \\
\hline
$\bar{q}c$&$1^{-}$&$D^*(2010)$& --- &---&---& --- & --- &--- \\
$\bar{q}c$&$2^{+}$&$D_2^*(2460)$& $(qq)c$ & $(3/2)^+$&$\Sigma_c^*(2520)$ & $(qq)[\bar{c}\bar{q}]$ & $1^{+}$ &${D}(2550)$ \\
$\bar{q}c$&$3^{-}$&$D_3^*(2750)$ & $(qq)c$ & $(3/2)^-$ & $\Sigma_c(2800)$ & $(qq)[\bar{c}\bar{q}]$ & --- & --- \\
 \hline
$\bar{s}c$&$0^{-}$&$D_s(1968)$& --- &---&---& --- & --- &--- \\
$\bar{s}c$&$1^{+}$&${D}_{s1}(2460)$& $[sq]c$ & $(1/2)^+$&$\Xi_c(2470)$ & $[sq][\bar{c}\bar{q}]$ & $0^{+}$ &$\bar{D}_{s0}^*(2317)$ \\
$\bar{s}c$&$2^{-}$&$D_{s2}(\sim2830)$? & $[sq]c$ & $(3/2)^-$ & $\Xi_c(2815)$ & $[sq][\bar{c}\bar{q}]$ & $1^{-}$ & --- \\
\hline
$\bar{s}c$&$1^{-}$&$D_s^*(2110)$& --- &---&---& --- & --- &--- \\
$\bar{s}c$&$2^{+}$&$D_{s2}^*(2573)$& $(sq)c$ & $(3/2)^+$&$\Xi_c^*(2645)$ & $(sq)[\bar{c}\bar{q}]$ & $1^{+}$ &${D}_{s1}(2536)$ \\
$\bar{s}c$&$3^{-}$&$D_{s3}^*(2860)$ & $(sq)c$ & $(1/2)^-$ & $\Xi_c(2930)$ & $(sq)[\bar{c}\bar{q}]$ & --- & --- \\
\hline
$\bar{c}s$&$1^{+}$&$\bar{D}_{s1}(\sim2700)$?& $[cs]s$ & $(1/2)^+$&$\Omega_c(2695)$ & $[cs][\bar{s}\bar{q}]$ & $0^{+}$ & ?? \\
\hline
$\bar{s}c$&$2^{+}$&$D_{s2}^*(\sim2750)$?& $(ss)c$ & $(3/2)^+$&$\Omega_c(2770)$ & $(ss)[\bar{c}\bar{s}]$ & $1^{+}$ & ?? 
\end{tabular}
\end{ruledtabular}
\end{table*}

\begin{table*}[ht]
\caption{\label{b} Same as Table~\ref{c} but for mesons containing bottom quarks from Ref.~\cite{Nielsen:2018uyn}. The quantum numbers $J^P=1^+,~J^P=0^+ $ and $J^P=2^-$ are assigned to the states $B_J(5732)$, $B_{J}^*(5840)$ and $B_J(5970)$, but their quantum numbers have not yet been determined. States with a question mark (?) are the predicted ones. The lowest meson of each family has no baryon or tetraquark partner and breaks the supersymmetry.}
 \begin{ruledtabular}
\begin{tabular}{ccc|ccc|ccc}
\multicolumn{3}{c|}{Meson} & \multicolumn{3}{c|}{Baryon} & \multicolumn{3}{c}{Tetraquark}\\
$q$-cont&$J^{P(C)}$ & Name & $q$-cont & $J^{P}$ &Name &$q$-cont &$J^{P(C)}$& Name \\
\hline
$\bar{q}b$&$0^-$&$\bar{B}(5280)$& --- &---&---& --- & --- &--- \\
$\bar{q}b$&$1^+$&$\bar{B}_1(5720)$& $[ud]b$ & $(1/2)^+$&$\Lambda_b(5620)$ & $[ud][\bar{b}\bar{q}]$ & $0^{+}$ &${B}_J(5732)$ \\
$\bar{q}b$&$2^{-}$&$\bar{B}_J(5970)$ & $[ud]b$ & $(3/2)^-$ & $\Lambda_b(5920)$ & $[ud][\bar{b}\bar{q}]$ &$1^{-}$ & --- \\
\hline
$\bar{b}q$&$0^{-}$&${B}(5280)$& --- &---&---& --- & --- &--- \\
$\bar{b}q$&$1^{+}$&${B}_1(5720)$& $[bq]q$ & $(1/2)^+$&$\Sigma_b(5815)$ & $[bq][\bar{u}\bar{d}]$ & $0^{+}$ &$\bar{B}_J(5732)$ \\
\hline
$\bar{q}b$&$1^{-}$&$B^*(5325)$& --- &---&---& --- & --- &--- \\
$\bar{q}b$&$2^{+}$&$B_2^*(5747)$& $(qq)b$ & $(3/2)^+$&$\Sigma_b^*(5835)$ & $(qq)[\bar{b}\bar{q}]$ & $1^{+}$ &${B}_J(5840)$ \\
  \hline
$\bar{s}b$&$0^{-}$&$B_s(5365)$& --- &---&---& --- & --- &--- \\
$\bar{s}b$&$1^{+}$&${B}_{s1}(5830)$& $[qs]b$ & $(1/2)^+$&$\Xi_b(5790)$ & $[qs][\bar{b}\bar{q}]$ & $0^{+}$ &$\bar{B}_{s0}^*(\sim5800)$? \\
\hline
$\bar{s}b$&$1^{-}$&$B_s^*(5415)$& --- &---&---& --- & --- &--- \\
$\bar{s}b$&$2^{+}$&$B_{s2}^*(5840)$& $(sq)b$ & $(3/2)^+$&$\Xi_b^*(5950)$ & $(sq)[\bar{b}\bar{q}]$ & $1^{+}$ &${B}_{s1}(\sim5900)$? \\
\hline
$\bar{b}s$&$1^{+}$&${B}_{s1}(\sim6000)$?& $[bs]s$ & $(1/2)^+$&$\Omega_b(6045)$ & $[bs][\bar{s}\bar{q}]$ & $0^{+}$ & ?? 
\end{tabular}
\end{ruledtabular}
\end{table*}

\end{document}